\RequirePackage{fix-cm}

\documentclass[twocolumn,epjc3]{svjour3}  

\RequirePackage[T1]{fontenc}

\smartqed  % flush right qed marks, e.g. at end of proof

\RequirePackage{graphicx}

\RequirePackage{mathptmx}      % use Times fonts if available on your TeX system

\RequirePackage{latexsym}
\RequirePackage[numbers,sort&compress]{natbib}
\RequirePackage[colorlinks,citecolor=blue,urlcolor=blue,linkcolor=blue]{hyperref}
\usepackage{enumitem} %para las sub-subsubsecciones Energy binning y Segmented detector
\usepackage{amssymb}

\begin{document}

\title{ANAIS--112 sensitivity in the search for dark matter annual modulation
}
%\subtitle{\\}


\author{I.~Coarasa\thanksref{addr1,addr2,e1}
        \and
        J.~Amar\'e\thanksref{addr1,addr2}
        \and
        S.~Cebri\'an\thanksref{addr1,addr2}
        \and
        C.~Cuesta\thanksref{addr1,addr2,addr4}
        \and
        E.~Garc\'ia\thanksref{addr1,addr2}
        \and
        M.~Mart\'inez\thanksref{addr1,addr2,addr3}
        \and
        M.A.~Oliv\'an\thanksref{addr1,addr5}
        \and
        Y.~Ortigoza\thanksref{addr1,addr2}
        \and
        A.~Ortiz de Sol\'orzano\thanksref{addr1,addr2}
        \and
        J.~Puimed\'on\thanksref{addr1,addr2}
        \and
        A.~Salinas\thanksref{addr1,addr2}
        \and
        M.L.~Sarsa\thanksref{addr1,addr2}       
        \and
        P.~Villar\thanksref{addr1,addr2}
        \and
        J.A.~Villar\thanksref{addr1,addr2,e2}
}

\thankstext{e1}{e-mail: icoarasa@unizar.es}
\thankstext{e2}{Deceased}

\institute{Grupo de F\'isica Nuclear y Astropart\'iculas, Universidad de Zaragoza, Calle Pedro Cerbuna 12, 50009 Zaragoza, Spain \label{addr1}
           \and
           Laboratorio Subterr\'aneo de Canfranc, Paseo de los Ayerbe s/n, 22880 Canfranc Estaci\'on, Huesca, Spain \label{addr2}
           \and
           Fundaci\'on Agencia Aragonesa para la Investigaci\'on y el Desarrollo, ARAID, Avenida de Ranillas 1--D, 50018 Zaragoza, Spain\label{addr3}
           \and
           \emph{Present Address:} Centro de Investigaciones Energ\'eticas, Medioambientales y
Tecnol\'ogicas, CIEMAT, 28040 Madrid, Spain\label{addr4}
           \and
           \emph{Present Address:} Centro de Investigaci\'on de Recursos y Consumos Energ\'eticos, CIRCE, 50018 Zaragoza, Spain\label{addr5}
}

\date{Received: date / Accepted: date}

\maketitle

\begin{abstract}

\begin{sloppypar}

The annual modulation measured by the ~DAMA/LIBRA~ experiment can be explained by the interaction of dark matter WIMPs in NaI(Tl) scintillator detectors. Other experiments, with different targets or techniques, exclude the region of parameters singled out by ~DAMA/LIBRA,~ but the comparison of their results relies on several hypotheses regarding the dark matter model. ANAIS--112 is a dark matter search with 112.5~kg of NaI(Tl) scintillators at the Canfranc Underground Laboratory (LSC) to test the ~DAMA/LIBRA~ result in a model independent way. We analyze its prospects in terms of the \textit{a priori} critical and detection limits of the experiment. A simple figure of merit has been obtained to compare the different experiments looking for the annual modulation observed by ~DAMA/LIBRA.~ We conclude that after 5~years of measurement, ANAIS--112 can detect the annual modulation in the $3\sigma$ region compatible with the DAMA/LIBRA result.

\keywords{Dark matter \and Annual modulation \and NaI(Tl) scintillator \and Critical limit \and Detection limit}

\end{sloppypar}

\end{abstract}

\section{Introduction}
\label{intro}

\begin{sloppypar}

The ANAIS experiment \cite{anais_taup2017,anais112performance_2018} is intended to search for dark matter annual modulation with ultrapure NaI(Tl) scintillators at the Canfranc Underground Laboratory (LSC) in Spain, in order to provide a model independent confirmation of the signal reported by the DAMA/LIBRA collaboration \cite{Bernabei2008,Bernabei20131,Bernabei2018} using the same target and technique. Projects like DM--Ice \cite{dmice2017_prd}, COSINE--100 \cite{cosine2017_epjc,cosine2018_epjc}, SABRE \cite{sabre2017_nima} and PICO--LON \cite{picolon_taup2015} also envisage the use of large
masses of NaI(Tl) for dark matter searches. Results obtained by other experiments with other target materials and techniques (like those from CDMS \cite{cdms2016_phrevl}, CRESST \cite{cresst2016_epjc}, EDELWEISS \cite{edelweiss2016_jcap}, KIMS \cite{kims2014_phrev}, LUX \cite{lux2014_phrevl}, PICO \cite{pico2016_phrev}, XENON \cite{xenon2015_phrevl} or DarkSide \cite{darkside2018_phrevl,darkside2018_prd} collaborations) have been ruling out for years the most plausible compatibility scenarios. Nevertheless, DAMA/LIBRA has accumulated up to now twenty annual cycles in the [2,6] keV$_{\textnormal{ee}}$ energy region (keV$_{\textnormal{ee}}$ for keV electron--equivalent) with $12.8\sigma$ statistical significance (phase and period fixed) \cite{Bernabei2008,Bernabei20131,Bernabei2018}. Moreover, DAMA/LIBRA--phase2 has been able to accumulate six annual cycles in the [1,6] keV$_{\textnormal{ee}}$ energy region with $9.5\sigma$ statistical significance because all the photomultipliers (PMTs) were replaced by a second generation PMTs Hamamatsu R6233MOD, with higher quantum efficiency and with lower background with respect to those used in phase1 \cite{Bernabei2018}.

The WIMP interaction counting rate experiences an annual modulation
as the result of the motion of the Earth around the Sun that can be approximated \cite{Freese_modulation88,Savage20091} by:
%suprimo la linea en blanco para igualar espacios antes y despues de la ecuación
\begin{equation}
\frac{dR}{dE_R}\left(E_R,t\right)\approx S_0\left(E_R\right)+S_m\left(E_R\right)\cdot{}\textnormal{cos}\left(2\pi\cdot{}\frac{t-t_0}{T}\right),
\label{eq:annualModulation}
\end{equation}
%suprimo la linea en blanco para evitar nuevo párrafo
where $R$ is the interaction rate, $E_R$ is the recoil energy, $t_0$ is the expected time of the maximum (or minimum, depending on the sign of $S_m$), 152.5 days after 1$^{st}$ January, and $T$ is the expected period of one year. The time--averaged differential rate is denoted by $S_0$, whereas the modulation amplitude is given by $S_m$ \cite{Savage20091}. The value of $S_m$ measured by DAMA/LIBRA is $0.0102\pm0.0008$ and $0.0105\pm0.0011$ cpd/kg/keV${_\textnormal{ee}}$ within [2,6] and [1,6] keV$_{\textnormal{ee}}$ intervals, respectively (cpd stands for \textit{counts per day}) \cite{Bernabei2018}. 

In this paper, we analyze the ANAIS--112 prospects in terms of the \textit{a priori} critical and detection limits of the experiment and a simple figure of merit has been obtained to compare the different experiments looking for the annual modulation observed by DAMA/LIBRA. The structure of the paper is as follows: section \ref{ANAISsetup} describes the ANAIS--112 experimental layout; section \ref{sec:2to6} focuses on the procedure to search for a modulation signal in the [2,6] keV$_{\textnormal{ee}}$ energy region, considering a single energy bin and afterwards the energy binning and segmented detector in nine modules; section \ref{sec:1to6} focuses on the [1,6] keV$_{\textnormal{ee}}$ energy region in view of the last DAMA/LIBRA--phase2 results. Finally, conclusions are presented in section \ref{concl}.

\section{The ANAIS--112 experiment}
\label{ANAISsetup}

\end{sloppypar}

ANAIS--112 consists of nine modules made by Alpha Spectra (AS), Inc. Colorado and then shipped to Spain along several years, arriving at LSC the first of them at the end of 2012 and the last by March, 2017. Each crystal is cylindrical (4.75$''$ diameter and 11.75$''$ length), with a mass of 12.5 kg. NaI(Tl) crystals were grown from selected ultrapure NaI powder and housed in OFE (Oxygen Free Electronic) copper; the encapsulation has a mylar window allowing low energy calibration. Two Hamamatsu R12669SEL2 PMTs were coupled through quartz windows to each crystal at LSC clean room. All PMTs have been screened for radiopurity using germanium detectors in Canfranc. The shielding for the experiment consists of 10 cm of archaeological lead, 20 cm of low activity lead, 40 cm of neutron moderator, an anti--radon box (continuously flushed with radon--free nitrogen) and an active muon veto system made up of plastic scintillators designed to cover top and sides of the whole ANAIS set--up. The hut housing the experiment is at the hall B of LSC under 2450 m.w.e.

The light output measured for all AS modules is at the level of $\sim$ 15 phe/keV$_{\textnormal{ee}}$ \cite{anais112performance_2018}, which is 1.5 times larger than that determined for the best DAMA/LIBRA detectors \cite{Bernabei2018}. This high light collection, possible thanks to the excellent crystal quality and the use of high quantum efficiency PMTs, has a direct impact in energy threshold. Triggering below 1~keV$_{\textnormal{ee}}$ is confirmed by the identification of bulk $^{22}$Na and $^{40}$K events at 0.9 and 3.2 keV$_{\textnormal{ee}}$, respectively, thanks to coincidences with the corresponding high energy photons following the electron capture decays to excited levels \cite{anais112performance_2018}. 

\begin{figure}
	\centering
		\includegraphics[width=0.5\textwidth]{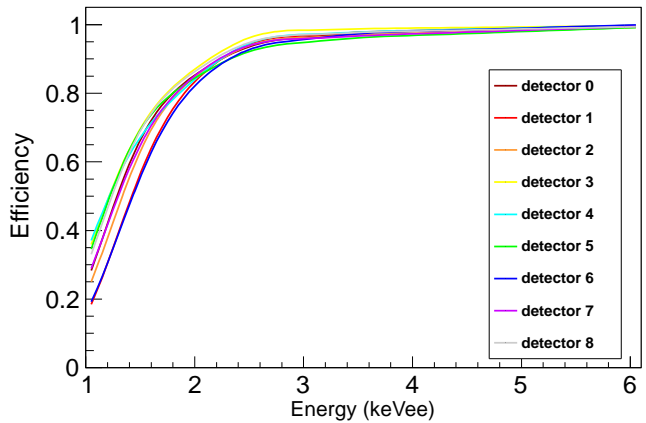}
	\caption{Total efficiency for every ANAIS--112 detector}\label{fig:efficiencies}
\end{figure}

To remove the PMT origin events, dominating the background below 10 keV$_{\textnormal{ee}}$, and then reach the 1 keV$_{\textnormal{ee}}$ threshold, specific filtering protocols for ANAIS--112 detectors have been designed. Multiparametric cuts based on the number of photoelectrons in the pulses, the temporal parameters of the pulses and the asymmetry in light sharing between PMTs are considered, and the corresponding acceptance efficiencies for such filters have been calculated. The trigger efficiency (probability that an event is triggered by the DAQ system) has been also considered \cite{anais112performance_2018}. The total efficiency for the selection of dark matter compatible events in every ANAIS--112 detector is shown in Fig.~\ref{fig:efficiencies}.

\begin{figure*}
	\centering
		\includegraphics[width=1.00\textwidth]{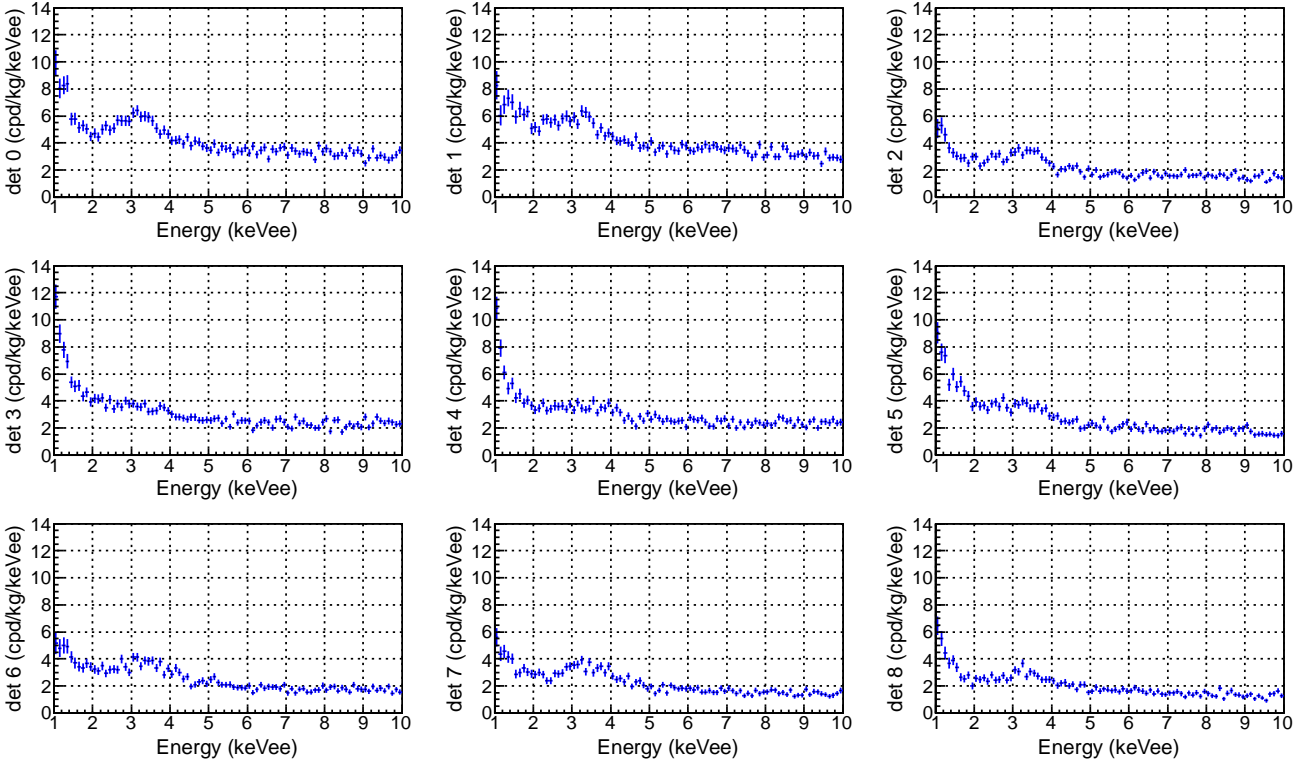}
	\caption{Anticoincidence energy spectrum measured at low energy after filtering and efficiency correction for each detector, corresponding to the $\sim$10\% of the first year of data unblinded}\label{fig:background3x3}
\end{figure*}

\begin{figure}
	\centering
		\includegraphics[width=0.5\textwidth]{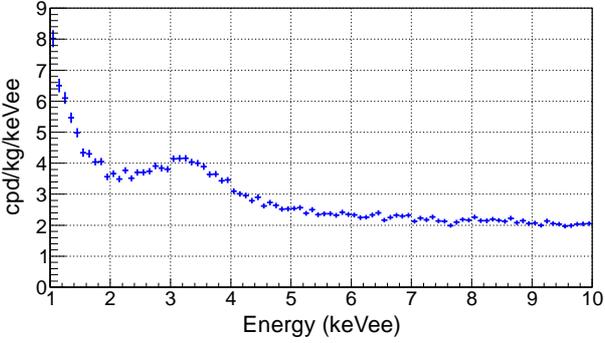}
	\caption{Total anticoincidence energy spectrum of ANAIS--112 at low energy corrected for efficiencies ($\sim$10\% of the first year of data unblinded)\label{fig:background}}
\end{figure}

ANAIS--112 dark matter run started on August 3, 2017. The first year of data taking finished on July 31, 2018, having accumulated 341.72 days of live time. The low energy data is blinded except for the multiplicity 2 events that we use to test the analysis procedures. On the other hand, $\sim$10\% of the first year of data has been unblinded for background assessment. To do this, one--day time bins have been selected randomly distributed throughout the whole year. The low energy spectra corresponding to the unblinded data for each of the detectors in anticoincidence (single hit events) after event selection and efficiency correction are shown in Fig.~\ref{fig:background3x3}. We also display in Fig.~\ref{fig:background} the total anticoincidence background (adding up all 9 detectors) below 10 keV$_{\textnormal{ee}}$ for the unblinded data.

At the region of interest, crystal bulk contamination is the dominant background source. Contributions from $^{40}$K, $^{210}$Pb (powder/crystal growing contaminants), $^{22}$Na and $^{3}$H (cosmogenics) are the most relevant \cite{epjc_bkg_d012,anais112bkgmodel_2018}. In almost all detectors the $^{40}$K peak at 3.2 keV$_{\textnormal{ee}}$ is clearly visible. The background level at 2 keV$_{\textnormal{ee}}$ ranges from 2 to 5 \linebreak cpd/kg/keV$_{\textnormal{ee}}$, depending on the detector, and then increases up to 5--8 cpd/kg/keV$_{\textnormal{ee}}$ at 1 keV$_{\textnormal{ee}}$. A detailed analysis of the background contributions can be found in \cite{anais112bkgmodel_2018}.

\section{Searching for a modulation signal in the [2,6] keV$_{\textnormal{ee}}$ energy interval} \label{sec:2to6}

A model independent way to check the DAMA/LIBRA result is looking for a signal not only with the same target and technique but also in the same region where DAMA/LIBRA finds it. First, we search for a modulation in the rate in a model independent way, i.e. without assumptions about the origin and characteristics of the signal other than the one--year period and the 152.5--days phase. To do this, we will evaluate in section \ref{ssec:2to6_noDM_hyp} the detection in [2,6] keV$_{\textnormal{ee}}$ of an annual modulation amplitude $b$ of the counting rate $B$
\begin{equation}
B(\tau)=a+b\textnormal{ cos }\tau,
\label{eq:uniformBkg}
\end{equation}
where $a$ is the mean annual rate and $\tau=2\pi(t-t_0)/T$, see Eq.~(\ref{eq:annualModulation}).
We will take the simplest approximation of only considering one bin (4~keV$_{\textnormal{ee}}$ wide) and the whole detection mass; afterwards we will take into account the energy binning and the segmentation of the 112.5~kg in 9 modules.

Finally, in section \ref{ssec:2to6_DM_hyp}, we will consider the particular case of a modulation induced by dark matter \cite{Cebrian2001339}.

\subsection{Model independent modulation} \label{ssec:2to6_noDM_hyp}

The test statistic \cite{Eadie_1971} to evaluate the null ($b=0$) and the alternative ($b\ne0$) hypotheses is the least squares estimator of the amplitude, $\hat{b}$, of expected value $E(\hat{b})=b$ and variance $var(\hat{b})$. Asymptotically, $\hat{b}$ follows a normal distribution.

The critical limit ($L_C$) is a threshold such that if $\hat{b}> L_C$, the signal is considered statistically significant. $L_C$ is defined from the distribution of $\hat{b}$ when there is no signal, $E(\hat{b})=0$. We use a one-tailed test because the amplitude measured by DAMA/LIBRA is positive. Then, for a confidence level $\alpha$, the probability of a false positive is $1-\alpha$:
\begin{equation}
P(\hat{b}\leq L_C\mid b=0)=\alpha
\end{equation}
The detection limit ($L_D$) is the modulation amplitude such that the outcome of its estimator $\hat{b}$ is greater than $L_C$ with $\beta$ probability:
\begin{equation}
P(\hat{b}>L_C\mid b=L_D)=\beta
\end{equation}

\subsubsection{A single energy bin} \label{sssec:2to6_nobinned}

A linear least--squares fit of Eq.~(\ref{eq:uniformBkg}) for $n$ time bins, where the dependent variable $B_i$ is the measured rate in the $i^{th}$ time bin $\tau_i$, $w_i=1/var(B_i)$ and the independent variable is $\textnormal{cos }\tau_i$, gives Eq. (6.12) on page 105 of Ref. \cite{Bevington200398}:
\begin{equation}
\hat{b}=\frac{\sum_l{w_lB_l\cdot{}\left(-\sum_i{w_i\textnormal{cos }\tau_i}+\textnormal{cos }\tau_l\cdot{}\sum_i{w_i}\right)}}{\sum_i{w_i}\cdot{}\sum_i{w_i\textnormal{cos}^2\tau_i}-\left(\sum_i{w_i\textnormal{cos }\tau_i}\right)^2}
\end{equation}
and, similarly, Eq. (6.22) on page 109 \cite{Bevington200398} translates into:
\begin{equation}
var(\hat{b})=\frac{\sum_i{w_i}}{\sum_i{w_i}\cdot{}\sum_i{w_i\textnormal{cos}^2\tau_i}-\left(\sum_i{w_i\textnormal{cos }\tau_i}\right)^2}
\label{eq:estimator_Variance}
\end{equation}

We can obtain a simple expression for $var(\hat{b})$. If $N_i$ is the number of observed events (Poisson distributed) and $\varepsilon$ is the fraction of true events remaining after the cuts to reject the noise and select true events:
\begin{equation}
B_i=\frac{N_i/\varepsilon}{\Delta E\cdot{}M\cdot{}\Delta t} \hspace{1.0 cm} var(B_i)=\frac{B_i/\varepsilon}{\Delta E\cdot{}M\cdot{}\Delta t}
\end{equation}
where $M$ is the total detection mass, $\Delta E$ and $\Delta t$ are the width of the energy and live time bins, respectively. Note that the number of observed events is divided twice by the efficiency in $var(B_i)$.

\begin{sloppypar}
If $b=0$, the expected value $E(B_i)=B$ is time independent. For $b\neq0$, $E(B_i)$ is nearly time independent if $b~\ll~a$, as is the case for the annual modulation measured by DAMA/LIBRA ($b\sim10^{-2}$ cpd/kg/keV$_{\textnormal{ee}}$) and the typical counting rates ($a\gtrsim 1$ cpd/kg/keV$_{\textnormal{ee}}$). Latter value guarantees also the normality of $\hat{b}$ for ANAIS--112, even for one--day time bins. Then:
\end{sloppypar}
\begin{equation}
var(B_i)\approx \frac{B/\varepsilon}{\Delta E\cdot{}M\cdot{}\Delta t}=\frac{1}{w}
\end{equation}

An unbiased sample of $\tau_i$ covering an integer number of periods guarantees that $\sum_{i}{\textnormal{cos }\tau_i}=~0$ and $\sum_{i}{\textnormal{cos}^2\tau_i}=~n\cdot{}\frac{1}{2}$. Therefore, Eq.~(\ref{eq:estimator_Variance}) is simplified because $\sum_{i}{w_i\textnormal{cos }\tau_i}\simeq$ \linebreak
$w\sum_{i}{\textnormal{cos }\tau_i}\simeq~0$ and
${\sum_{i}{w_i\textnormal{cos}^2\tau_i}~\simeq~w\sum_{i}{\textnormal{cos}^2\tau_i}~\simeq~w\cdot{}n\cdot{}\frac{1}{2}}$; then, $var(\hat{b})$ is
\begin{equation}
var(\hat{b})=\frac{2\cdot{}B}{\Delta E\cdot{}M\cdot{}T_M\cdot{}\varepsilon} \hspace{1.0cm} \left(b\ll a\right)
\label{eq:bVariance}
\end{equation}
with $T_M=n\cdot{}\Delta t$ the measurement time.

$L_C$ and $L_D$ are proportional to the standard deviation $\sigma(\hat{b})=\sqrt{var({\hat{b})}}$, which can be used as a figure of merit to compare the different experiments looking for the annual modulation observed by DAMA/LIBRA.
\begin{equation}
FOM=\sqrt{{\frac{2\cdot{}B}{\Delta E\cdot{}M\cdot{}T_M\cdot{}\varepsilon}}}
\label{eq:fom}
\end{equation}

We have estimated $B$ and $\varepsilon$ of the nine modules for the $\sim$10\% unblinded data. The background of all modules in [2,6] keV$_{\textnormal{ee}}$ is listed in the 2$^{nd}$ column of Table~\ref{table:detectorsBkg} and the cut efficiencies, $\varepsilon$, are shown in Fig.~\ref{fig:efficiencies}. These are comparable in the [2,6] keV$_{\textnormal{ee}}$ interval, with average $\varepsilon=0.97$.

\begin{table}
\caption{Measured $\sim$10\% unblinded background (Fig.~\ref{fig:background}) in the [2,6] keV$_{\textnormal{ee}}$ energy interval for all modules after filtering and efficiency correction (Fig.~\ref{fig:efficiencies}) have been applied ($2^{nd}$ column). Considering the energy binning, the relevant quantity is $\left\langle B/\varepsilon\right\rangle$ ($3^{rd}$ column), where the background is divided twice by the efficiency, see section \ref{sssec:2to6_binned}. The average values for ANAIS--112 are listed in the last row. \label{table:detectorsBkg}}
\begin{tabular*}{\columnwidth}{@{\extracolsep{\fill}}ccc@{}}
\hline
 & $B$ & $\left\langle B/\varepsilon\right\rangle$ \\
Module & (cpd/kg/keV$_{\textnormal{ee}}$) & (cpd/kg/keV$_{\textnormal{ee}}$)\\ 
\hline
 D0 &  4.58$\pm$0.05 & 4.74$\pm$0.05 \\
 D1 &  4.66$\pm$0.05 & 4.82$\pm$0.05 \\
 D2 &  2.44$\pm$0.04 & 2.54$\pm$0.04 \\
 D3 &  3.16$\pm$0.04 & 3.24$\pm$0.04 \\
 D4 &  3.12$\pm$0.04 & 3.22$\pm$0.04 \\
 D5 &  2.96$\pm$0.04 & 3.11$\pm$0.04 \\
 D6 &  2.90$\pm$0.04 & 3.02$\pm$0.04 \\
 D7 &  2.61$\pm$0.04 & 2.72$\pm$0.04 \\
 D8 &  2.29$\pm$0.04 & 2.37$\pm$0.04 \\
 ANAIS--112 & 3.19$\pm$0.01 & 3.31$\pm$0.01 \\
\hline
\end{tabular*}
\end{table}

The usual results of the experiments looking for dark matter are exclusion plots (upper limits) at 90\% C.L. in the plane cross section WIMP--nucleon versus WIMP mass \cite{Savage20091}.
By definition of $L_D$, $L_D\simeq 2L_C$ if $var(\hat{b})\simeq var(\hat{b}\mid b=0)$ and both are set to the same C.L. (see Fig. 2 of Ref. \cite{currie_1968}). ANAIS--112 fulfills the former condition because $b\ll a$, see Eq.~(\ref{eq:bVariance}). Under the same conditions as above, any upper limit, $L_U$, satisfies $L_U\leq L_D$ \cite{currie_1968}. Furthermore, $L_C\leq L_U$ (both to the same C.L.) if the outcome of $\hat{b}$ is $\geq0$. If $\hat{b}<0$, it should be $\mid\hat{b}\mid\sim\sigma(\hat{b})$ because if $\hat{b}\ll-\sigma(\hat{b})$, it would imply a negative modulation, opposite to the observed by DAMA/LIBRA. Briefly, any $L_U$ given by ANAIS--112 will be less than $L_D$, likely greater than $L_C$ or, at least, not much smaller than $L_C$.

Therefore, in order to compare properly the expectations of ANAIS--112 with other experiments, we also chose the 90\% C.L. for $L_C$ and $L_D$. Then, $L_C=1.28\cdot{}\sigma(\hat{b})$ and $L_D=2L_C$. Using Eq.~(\ref{eq:bVariance}) with $B=3.19 \pm 0.01$~cpd/kg/keV$_{\textnormal{ee}}$ (average background, Table \ref{table:detectorsBkg}), $\Delta E=4$ keV$_{\textnormal{ee}}$, $M=112.5$ kg, $T_M=5$ years and $\varepsilon=0.97$:
\begin{equation}
L_D=(7.24\pm 0.02)\cdot 10^{-3} \textnormal{ cpd/kg/keV}_{\textnormal {ee}} \hspace{0.3cm}\left(90\%\textnormal{ C.L.}\right)
\label{eq:aSingleEnergyBin}
\end{equation}
\begin{sloppypar}
\noindent
that is less than the DAMA/LIBRA signal. Then, ANAIS--112 can detect it. Furthermore, if the estimator of the DAMA/LIBRA signal is normal, with mean and standard deviation 0.0102 and 0.0008 cpd/kg/keV${_\textnormal{ee}}$, respectively, about 0.01\% of the probability distribution is below our central value for $L_D$.
\end{sloppypar}

It is worth noting that, assuming a background linearly decreasing with time as an approximation of the decay of the long--lived $^{210}$Pb and $^{3}$H \cite{epjc_bkg_d012} during data taking, the obtained $L_D$ is very similar to the one obtained assuming a constant background \cite{TFM_ivan}. The contribution of $^{210}$Pb ($^{3}$H) in the [2,6] keV$_{\textnormal{ee}}$ has been estimated for the first year of data taking as 1.246 (0.826) cpd/kg/keV$_{\textnormal{ee}}$ \cite{anais112bkgmodel_2018}. Therefore, adding a linear term to Eq.~(\ref{eq:uniformBkg}), a three parameter linear least--squares fit \cite{Bevington200398} can be carried out and \linebreak
$L_D=7.20\cdot{}10^{-3}$ cpd/kg/keV$_{\textnormal{ee}}$ is obtained.

\subsubsection{Energy binning and segmented detector} \label{sssec:2to6_binned}

A more accurate $L_C$ and $L_D$ value can be obtained taking into account the energy binning and the background and efficiency differences among the modules of ANAIS--112 (segmented detector). In addition, the energy binning and the segmented detector in nine modules should be considered to obtain the possible energy dependence of the modulation amplitude $b(E)$.

\begin{enumerate}[label=($\alph*$)]
	\item \textit{Energy binning}
\end{enumerate}

The average annual modulation amplitude in the [2,6] keV$_{\textnormal{ee}}$ interval is
\begin{equation}
b=\frac{1}{\Delta E}\int_{E_1}^{E_1+\Delta E}b\left(E\right)dE,
\label{eq:bAmplitude}
\end{equation}
where $E_1=2$ and $\Delta E=4$~keV$_{\textnormal{ee}}$. Then, for $N$ bins the $j^{th}$ modulation amplitude in $\left[E_j,E_{j+1}\right]$ ($j=1,2,\cdots,N$) is
\begin{equation}
b_j=\frac{1}{\Delta E_j}\int_{E_j}^{E_{j+1}}b\left(E\right)dE,
\label{eq:bjAmplitude}
\end{equation}
being $\Delta E_j=E_{j+1}-E_{j}$. According to Eq.~(\ref{eq:bVariance})
\begin{equation}
var(\hat{b}_j)=\frac{2\cdot{}B_j}{\Delta E_j\cdot{}M\cdot{}T_M\cdot{}\varepsilon_j}
\label{eq:bjVariance}
\end{equation}
where $B_j$ and $\varepsilon_j$ are the background and the efficiency in the $j^{th}$ bin, respectively.
If all the bins are of equal width $\Delta E_j=\Delta E/N\equiv \delta E$, then
\begin{equation}
b=\frac{1}{N\cdot{}\delta E}\sum_{j=1}^{N}{\int_{E_j}^{E_{j+1}}b\left(E\right)dE}=\frac{1}{N}\cdot{}\sum_{j=1}^{N}{b_j}
\end{equation}
so that $b$ is the arithmetic mean of $b_j$.
For $N=40$ ($\delta E=0.1$~keV$_\textnormal{ee}$) and $M=112.5$~kg, $\hat{b}_j$'s are also virtually normal variables for one--day time bins. When the $\hat{b}_j$'s are statistically independent
\begin{equation}
var(\hat{b})=\frac{1}{N^2}\cdot{}\sum_{j=1}^{N}var(\hat{b}_j)=\frac{2\cdot{}\left\langle B/\varepsilon \right\rangle}{\Delta E\cdot{}M\cdot{}T_M}
\label{eq:bVarianceBinning}
\end{equation}
with $\left\langle B/\varepsilon \right\rangle=(1/N)\cdot{}\sum_{j=1}^{N}{B_j/\varepsilon_j}$. The detection limit obtained is identical to Eq.~(\ref{eq:aSingleEnergyBin}) because $B(E)/\varepsilon(E)$ is nearly energy independent in [2,6] keV$_\textnormal{ee}$.

\begin{enumerate}[resume, label=($\alph*$)]
	\item \textit{Segmented detector}
\end{enumerate}

We consider now the data of each module. According to Eq.~(\ref{eq:bjVariance}), the variance of the estimator of the modulation amplitude in the $j^{th}$ energy bin of the module $k$ ($k=1,2,\cdots,9$) is:
\begin{equation}
var(\hat{b}_j^k)=\frac{2\cdot{}B_j^k}{\delta E\cdot{}m\cdot{}T_M\cdot{}\varepsilon_j^k}
\end{equation}
where $m=12.5$ kg is the mass of one module and $B_j^k$ and $\varepsilon_j^k$ are the background and the efficiency in the $j^{th}$ energy bin of the module $k$. Now, $\hat{b}_j^k$'s are virtually normal variables for one--week time bins. Thus, the variance of the estimator of $b$ in the module $k$ is:
\begin{equation}
var(\hat{b}^k)=\frac{2\cdot{}\left\langle B/\varepsilon \right\rangle^k}{\Delta E\cdot{}m\cdot{}T_M}
\end{equation}
with $\left\langle B/\varepsilon \right\rangle^k=(1/N)\cdot{}\sum_{j=1}^{N}{B_j^k/\varepsilon_j^k}$. The estimator $\hat{b}$ with the nine modules is the weighted mean of the nine $\hat{b}^k$ and its variance is:
\begin{eqnarray}
var(\hat{b}) &=& \left(\sum_{k=1}^{9}{\frac{1}{var(\hat{b}^k)}}\right)^{-1} \nonumber \\
             &=& \frac{2}{\Delta E\cdot{}m\cdot{}T_M}\cdot{}\left(\sum_{k=1}^{9}{\frac{1}{\left\langle B/\varepsilon\right\rangle^k}}\right)^{-1}
\label{eq:9detectorsVariance}
\end{eqnarray}

\begin{sloppypar}
According to the $3^{rd}$ column of Table \ref{table:detectorsBkg}, \linebreak $L_D=(7.07\pm0.02)\cdot{}10^{-3}$ cpd/kg/keV$_\textnormal{ee}$. The approximation of Eq.~(\ref{eq:aSingleEnergyBin}) is very close to this value because the nine values ${\left\langle B/\varepsilon\right\rangle}^k$ are close to ${\left\langle B/\varepsilon\right\rangle}$ and $B(E)/\varepsilon(E)$ is nearly constant (energy independent) in [2,6] keV$_\textnormal{ee}$.
\end{sloppypar}

\subsection{Dark matter hypothesis} \label{ssec:2to6_DM_hyp}

This hypothesis means that the possible modulation has to be compatible with the energy dependence of the modulation amplitude, $b(E;\sigma,M_{WIMP})$ \cite{primack_seckel_sadoulet}, where ~$\sigma$ is the WIMP--nucleon cross section and $M_{WIMP}$ the WIMP mass (we follow the most common framework for dark matter detection).
We take the differential rate from \cite{Savage20091}, the local dark matter density $\rho~=~0.3$~GeV/cm$^3$, the most probable WIMP velocity $v_0~=~220$~km/s and the escape velocity $v_{esc}~=~650$~km/s.
We consider the spin--independent WIMP interaction, using the Helm nuclear form factor \cite{Lewin199687} and $Q_{Na}~=~0.30$ and $Q_I~=~0.09$ for the sodium and iodine quenching factors to transform the nuclear recoil energy into electron equivalent one, respectively \cite{Savage20091}.
The resolution of ANAIS--112 is estimated in [1,6]~keV$_\textnormal{ee}$ as $\Gamma=0.890 \sqrt{E(\textnormal{keV}_\textnormal{ee})}-0.188$, where $\Gamma$ is the full width at half maximum \cite{anais112performance_2018}.
The Earth velocity \cite{Bernabei2013370} is given by
\begin{equation}
v_E(t)=232+15\cdot{}\textnormal{ cos}\left(2\pi\cdot{}\frac{t-152.5}{365.25}\right) \textnormal{ km/s,}
\label{eq:EarthVelocity}
\end{equation}
with the maximum value at $t=152.5$ days (2$^{nd}$ June). 

The test statistic in this case is the maximum likelihood ratio, which we already used in a more general context \cite{Cebrian2001339}. It is asymptotically equivalent to test the difference between the $\chi^2_{min}$ of the null ($\sigma=0$) and alternative ($\sigma\ne0$) hypotheses \cite{Eadie_1971}. This equivalence is easily satisfied for 0.1~keV$_\textnormal{ee}$ energy bins, see section \ref{sssec:2to6_binned}. The minimum of
\begin{equation}
\chi^2(\sigma,M_{WIMP})=\sum_{j}{\frac{\left(\hat{b}_j-b_j(\sigma,M_{WIMP})\right)^2}{var(\hat{b}_j)}},
\label{eq:chi2}
\end{equation}
has to be evaluated for $\sigma=0$ and $\sigma\ne0$. If $\sigma=0$, the quantity
\begin{equation}
\Delta\chi^2=\chi^2(\sigma=0,M_{WIMP})_{min}-\chi^2(\sigma,M_{WIMP})_{min}
\end{equation}
is distributed as a $\chi^2_\nu$ variable with $\nu=2$ degrees of freedom. $L_C$ at 90\% C.L. is such that $P(\chi^2_2\leq L_C)=0.9$, $L_C=4.61$.
On the other hand, if $\sigma\ne0$, $\Delta\chi^2$ is a non--central $\chi'^2_{(\nu,\lambda)}$ with $\nu=1$ degree of freedom, expected value
\begin{equation}
\left\langle\Delta\chi^2\right\rangle=\frac{1}{2}\cdot{}\sum_{j}{\frac{b_j^2\cdot{}\Delta E_j\cdot{}\varepsilon_j}{B_j}\cdot{}M\cdot{}T_M}+2,
\label{eq:likelihoodRatio}
\end{equation}
(see Ref. \cite {Cebrian2001339}) and non--central parameter $\lambda=\left\langle\Delta\chi^2\right\rangle-1$. The detection limit at 90\% C.L. is defined by $P(\chi'^2_{(1,\lambda)}>L_C)=0.9$, that holds when $\left\langle\Delta\chi^2\right\rangle=12.8$.

The segmented detector can be incorporated to the test, obtaining
\begin{equation}
\left\langle\Delta\chi^2\right\rangle=\frac{1}{2}\cdot{}\sum_{j,k}{\frac{(b_j^k)^2\cdot{}\Delta E_j\cdot{}\varepsilon_j^k}{B_j^k}\cdot{}m\cdot{}T_M}+2.
\label{eq:likelihoodRatioSpatialBinning}
\end{equation}

The value of $\left\langle\Delta\chi^2\right\rangle=12.8$ in Eq.~(\ref{eq:likelihoodRatioSpatialBinning}) defines our detection limit as an implicit function $\sigma\left(M_{WIMP}\right)$ because \linebreak $b_j^k\left(\sigma,M_{WIMP}\right)$ and the other variables of Eq.~(\ref{eq:likelihoodRatioSpatialBinning}) are experimental data. The same is valid for Eq.~(\ref{eq:likelihoodRatio}).

\begin{figure}
	\includegraphics[width=0.5\textwidth]{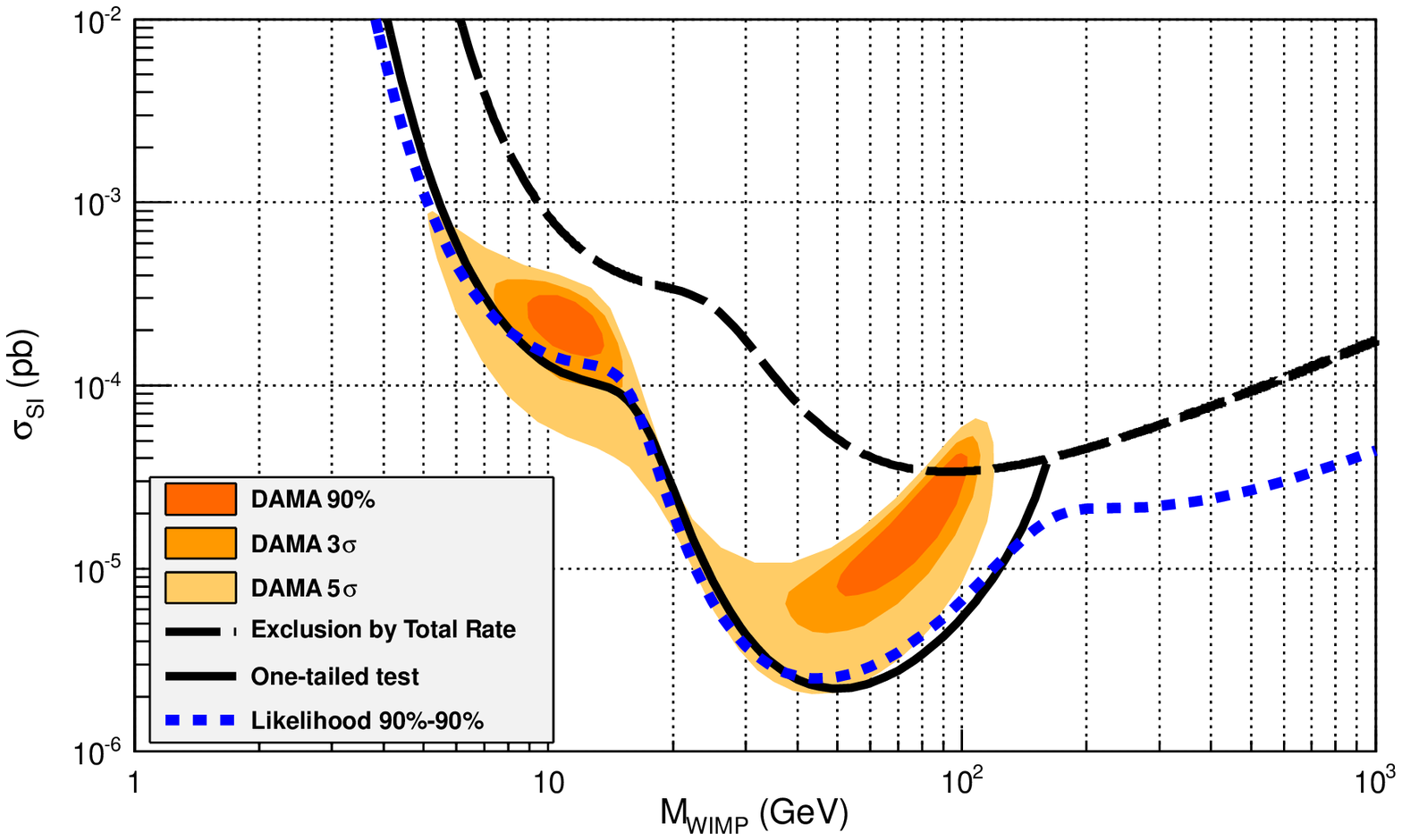}
	\caption{Result of the maximum likelihood test ratio for the detection limit in the [2,6] keV$_{\textnormal{ee}}$ window at 90\% C.L. (when critical limit is at 90\% C.L.) with 40 energy bins and segmented detector of ANAIS--112 after 5 years of measurement (short dashed blue line). The exclusion by total rate for the spin--independent WIMP--nucleon cross section of ANAIS--112 is the long dashed black line. DAMA/LIBRA regions at 90\%, 3$\sigma$ and 5$\sigma$ are also shown \cite{Savage20091}.
The detection limit (section \ref{sssec:2to6_nobinned}) is the solid black line \label{fig:OneTailedAndLikelihood}}
\end{figure}

The detection limit under the dark matter hypothesis is shown in the Fig.~\ref{fig:OneTailedAndLikelihood}, taking the background shown in the Fig.~\ref{fig:background} and an exposure of $M\cdot{}T_M=112.5$ kg$\times 5$ years. \linebreak
ANAIS--112 can detect the annual modulation of the interaction rate of WIMPs with Na or I in almost all the $3\sigma$ DAMA/LIBRA region \cite{Savage20091}. The region above the long \linebreak dashed black line of Fig.~\ref{fig:OneTailedAndLikelihood} is excluded at 90\% C.L. because the dark matter rate in [2,6] keV$_{\textnormal{ee}}$ is greater than the observed one. The line has been calculated using the "binned Poisson" method \cite{Savage20091} for the binned data in [2,6] keV$_{\textnormal{ee}}$.

The one--tailed $L_D$ of Eq.~(\ref{eq:aSingleEnergyBin}), deduced from the figure of merit Eq.~(\ref{eq:fom}), can be translated to the $(\sigma_{SI},M_{WIMP})$ plane, see the solid black line of the Fig.~\ref{fig:OneTailedAndLikelihood}. For ANAIS--112, it is numerically equivalent to the maximum likelihood ratio test under the dark matter hypothesis. For $M_{WIMP}>180$~GeV the modulation amplitude is negative in the [2,6] keV$_{\textnormal{ee}}$ energy interval, a result non considered in the one--tailed test because it is opposite to the DAMA/LIBRA signal.

\section{ANAIS--112 in the [1,6] keV$_{\textnormal{ee}}$ energy interval} \label{sec:1to6}

The case of a single energy bin is not a good approximation because $B(E)/\varepsilon(E)$ changes steeply below 2~keV$_\textnormal{ee}$ (Fig.~\ref{fig:background}).
In order to estimate $L_C$, a one--tailed test is carried out again, $L_C=1.28\cdot{}\sigma(\hat{b})$ and $L_D=2L_C$.

\begin{sloppypar}
For $N=50$ ($\delta E=0.1$~keV$_\textnormal{ee}$), $\hat{b}_j$ and $\hat{b}_j^k$ are normal variables as in section~\ref{sssec:2to6_binned}.
Taking $var(\hat{b})$ of Eq.~(\ref{eq:bVarianceBinning}) and $\left\langle B/\varepsilon\right\rangle=4.73 \pm 0.02$~cpd/kg/keV$_\textnormal{ee}$ (last row of Table~\ref{table:detectorsBkg[1,6]}), $L_D$ at 90\% C.L. (when $L_C$ is at 90\%) of ANAIS--112 after 5 years is:
\end{sloppypar}
\begin{equation}
L_D=(7.77\pm 0.01)\cdot 10^{-3} \textnormal{ cpd/kg/keV}_{\textnormal {ee}} \hspace{0.3cm}\left(90\%\textnormal{ C.L.}\right)
\label{eq:energyBinning[1,6]}
\end{equation}
\begin{sloppypar}
\noindent
that is less than the DAMA/LIBRA signal. Then, ANAIS--112 can detect it. Furthermore, if the estimator of the DAMA/LIBRA signal is normal, with mean and standard deviation 0.0105 and 0.0011 cpd/kg/keV${_\textnormal{ee}}$, respectively, less than 0.7\% of the probability distribution is below our central value for $L_D$.
\end{sloppypar}

\begin{table}
\caption{$\left\langle B/\varepsilon\right\rangle$ calculated from measured $\sim$10\% unblinded background in the [1,6] keV$_{\textnormal{ee}}$ energy interval for all modules after filtering and efficiency correction have been applied. The average values for ANAIS--112 are listed in the last row. \label{table:detectorsBkg[1,6]}}
\centering
\begin{tabular*}{5cm}{@{\extracolsep{\fill}}cc@{}}
\hline
 & $\left\langle B/\varepsilon\right\rangle$ \\
Module & (cpd/kg/keV$_{\textnormal{ee}}$)\\ 
\hline
 D0 & 6.42$\pm$0.06 \\
 D1 & 7.04$\pm$0.06 \\
 D2 & 3.59$\pm$0.04 \\
 D3 & 4.91$\pm$0.05 \\
 D4 & 4.60$\pm$0.05 \\
 D5 & 4.58$\pm$0.05 \\
 D6 & 4.48$\pm$0.05 \\
 D7 & 3.67$\pm$0.04 \\
 D8 & 3.29$\pm$0.04 \\
 ANAIS--112 & 4.73$\pm$0.02 \\
\hline
\end{tabular*}
\end{table}

Taking each module separately, according to the Table \ref{table:detectorsBkg[1,6]} and the Eq.~(\ref{eq:9detectorsVariance}), $L_D=(7.55\pm 0.02)\cdot{}10^{-3}$ cpd/kg/keV$_\textnormal{ee}$, very close to Eq.~(\ref{eq:energyBinning[1,6]}) because the nine values ${\left\langle B/\varepsilon\right\rangle}^k$ are close to ${\left\langle B/\varepsilon\right\rangle}$.

\begin{figure}
	\includegraphics[width=0.5\textwidth]{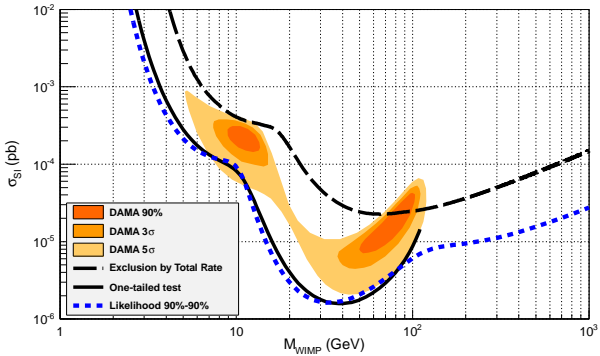}
	\caption{Result of the maximum likelihood test ratio for the detection limit in the [1,6] keV$_{\textnormal{ee}}$ window at 90\% C.L. (when critical limit is at 90\% C.L.) with 50 energy bins and segmented detector of ANAIS--112 after 5 years of measurement (short dashed blue line). The exclusion by total rate for the spin--independent WIMP--nucleon cross section of ANAIS--112 is the long dashed black line. DAMA/LIBRA regions at 90\%, 3$\sigma$ and 5$\sigma$ are also shown \cite{Savage20091}. The detection limit is the solid black line \label{fig:OneTailedAndLikelihood16}}
\end{figure}

\subsection{Dark matter hypothesis}

\begin{sloppypar}
The detection limit of ANAIS--112 at 90\% C.L. (for $L_C$ at 90\% C.L.) under the dark matter hypothesis is shown in the Fig.~\ref{fig:OneTailedAndLikelihood16}, taking the same exposure used in [2,6] keV$_\textnormal{ee}$ (Fig.~\ref{fig:OneTailedAndLikelihood}). The region of detection is now bigger for $M_{WIMP}<50$~GeV, because the background increasing is compensated by a higher signal below 2~keV$_\textnormal{ee}$ (see Eq.~(\ref{eq:likelihoodRatio}) and Eq.~(\ref{eq:likelihoodRatioSpatialBinning})). For $M_{WIMP}>110$~GeV the modulation amplitude is negative in the [1,6] keV$_{\textnormal{ee}}$ energy interval, a result non considered in the one--tailed test because it is opposite to the ~DAMA/LIBRA~ signal. It is worth noting that the DAMA/LIBRA region displayed in the plot is not updated for the new DAMA/LIBRA data between [1--6] keV$_\textnormal{ee}$, and serves only as reference. According to Ref. \cite{Baum2019789}, the regions of masses shift to the left in the $(\sigma_{SI},M_{WIMP})$ plane (from $\sim$10~GeV to $\sim$8~GeV for low mass and from $\sim$70~GeV to $\sim$54~GeV for high mass) but, in any case, the regions singled out by DAMA/LIBRA are above our detection limit.
\end{sloppypar}

\section{Conclusions}
\label{concl}

\begin{sloppypar}
We have estimated the detection limit at 90\% C.L., when the critical limit is at 90\% C.L., of ANAIS--112 for the annual modulation observed by DAMA/LIBRA. It is based on the measured background following the unblinding of $\sim$10\% of the first year of data of the nine modules D0 to D8. In the two considered scenarios (the [2,6] keV$_\textnormal{ee}$ and the [1,6] keV$_\textnormal{ee}$), we conclude that after 5~years of measurement, ANAIS--112 can detect the annual modulation in the $3\sigma$ region compatible with the DAMA/LIBRA result. The sensitivity in [2,6] keV$_\textnormal{ee}$ is very similar to that obtained in previous paper \cite{ivan_PLBarxiv}, where the background estimation was based on the measured activity of the six modules D0 to D5. On the other hand, the sensitivity in [1,6] keV$_\textnormal{ee}$ is now much better due to the improvements introduced in the efficiency estimate below 2 keV$_\textnormal{ee}$.
\end{sloppypar}

We give a simple figure of merit that gives good estimates of $L_C$ and $L_D$ if the ratio $B(E)/\varepsilon(E)$ is nearly constant (energy and detector independent), as it is our case within [2,6] keV$_\textnormal{ee}$. Furthermore, in order to compare the sensitivity of different experiments looking for the annual modulation, several approaches depending on the available information are also provided.

\begin{acknowledgements}

\begin{sloppypar}
Professor J.A. Villar passed away in August, 2017. Deeply in sorrow, we all thank his dedicated work and kindness. This work has been supported by the Spanish Ministerio de Econom\'ia y Competitividad and the European Regional Development Fund (MINECO-FEDER) (FPA2014-55986-P and FPA2017-83133-P), the Consolider-Ingenio 2010 Programme under grants MULTIDARK CSD2009-00064 and CPAN CSD2007-00042, and the Gobierno de Arag\'on and the European Social Fund (Group in Nuclear and Astroparticle Physics, ARAID Foundation and I. Coarasa predoctoral grant). We also acknowledge LSC and GIFNA staff for their support.
\end{sloppypar}

\end{acknowledgements}

\end{document}